\documentclass[twocolumn]{aa} 
\usepackage{color}

\usepackage{graphicx}
\usepackage{txfonts}

\usepackage{natbib}
\bibpunct{(}{)}{;}{a}{}{,} 

\begin{document}



\title{Nucleosynthesis of elements in gamma-ray burst engines}
\author{Agnieszka Janiuk\inst{1}
}

\institute{Center for Theoretical Physics, 
Polish Academy of Sciences, Al. Lotnikow 32/46, 02-668 Warsaw, Poland \\
\email{agnes@cft.edu.pl}
}

   \date{Received ...; accepted ...}
  \abstract
{}
{
We consider the gamma-ray burst (GRB) central engine that is powered by
the collapse of a massive rotating star or 
compact binary merger. 
The engine is a hot and dense accretion disk, which is composed of free nucleons, electron-positron pairs, and Helium, and cooled by neutrino emission.
A significant number density of neutrons in the inner disk body 
will conditions for neutron rich plasma
in the GRB outflows or jets. 
Helium is synthesized in the inner disk if the accretion rate is large, and heavy nuclei are also formed 
in the outer disk at distances above 150-250 $r_{g}$ from the black hole.
We study the process of nucleosynthesis in the GRB engine, depending on its 
physical properties.
}
   {
The GRB central engine is hydrodynamically modelled 
in the frame of a dense and hot disk, which
accretes with a high rate (up to 1 Solar mass per second) 
onto a maximally spinning, stellar mass black hole.
The synthesis of heavy nuclei up to Germanium and Gallium is then
followed by the nuclear reaction network.
}
{
The accretion at high rate onto a Kerr black hole 
feeds the engine activity and establishes conditions
for efficient synthesis of heavy nuclei in the disk. 
These processes may have important observational implications for
the jet deceleration process and heavy elements observed in the spectra of
GRB afterglows.
}
{}
  \authorrunning{A. Janiuk}
  \titlerunning{Nucleosynthesis in GRBs}
   \maketitle

\section{Introduction}

Gamma-ray bursts are transient sources of extreme brightness
observed on the sky with isotropic distribution. The extreme energetics of the 
observed bursts, detectable from a cosmological distance, implies that 
they must be connected with a gravitational potential energy released by 
accretion onto a compact star. The short bursts 
($T_{90}$ < 2 s) are likely related with mergers of binary compact star systems, such as 
binary neutron stars (NS-NS) or black hole-neutron star (BH-NS) binaries, while the
long duration events are believed to originate in collapsing 
massive stars. 
The typical mass of an accreting, newly born black hole is therefore on 
the order of a few Solar masses, while the accretion rates might exceed 
1 $M_{\odot} s^{-1}$.

In the collapsar model, the black hole 
surrounded by the part of the fallback stellar 
envelope helps launching relativistic jets 
\citep{1993ApJ...405..273W, 1999ApJ...524..262M}. 
These polar jets give rise to
gamma rays, which are produced far away from the 'engine' in the circumstellar
region 
(see, e.g., the reviews by \citealt{2004IJMPA..19.2385Z}, 
\citealt{2004RvMP...76.1143P}).

The process of nucleosynthesis of heavy elements 
in the central engines of gamma ray bursts has recently been studied 
in a number of works. 
The evolution of abundances calculated by  \citet{2013ApJ...778....8B}
have shown the synthesis of rare elements, such as $^{31}P$, $^{39}K$, 
$^{43}Sc$, and $^{35}Cl$ and other uncommon isotopes. 
These elements, which are produced in the simulations 
at outer parts of low $\dot M$ accretion disks (i.e. 0.001-0.01 
$M_{\odot} s^{-1}$), have been discovered in the emission lines of some
long GRB X-ray afterglows; however, they are yet to be confirmed by 
future observations.

In this article, we consider the nucleosynthesis of elements in the 
accretion disk itself for higher initial accretion rates, as is appropriate 
for type I collapsars
or neutron star mergers and short GRBs.
We account for the neutrino opacities and nuclear equation of state in 
dense matter. 
For the accretion disk model, we use the equation of state 
that is introduced in \citet{2007ApJ...664.1011J}
and subsequently adopted in \citet{2010A&A...509A..55J} to describe the disk
around a Kerr black hole with an arbitrary spin, $a$.
Contrary to previous work, which analytically accounted for the
total pressure that consists of gas, radiation, and 
completely degenerate electrons
\citep{1999ApJ...518..356P}
with subsequent addition of the neutrino pressure \citep{2002ApJ...579..706D},
we compute the equation of state (EOS) of matter numerically from the nuclear reaction balance. 
Our approach allows for a partial degeneracy of all the species, a 
partial trapping of neutrinos, and Kerr black hole solutions. 

We calculate the profiles of density and temperature, as well as the electron fraction
in the converged steady-state model of an accreting torus in
the GRB engine. We then follow the nucleosynthesis 
process, and we determine the abundances of the heavy elements isotopes.

\section{Neutronization in the hyperdense matter}

The central engines of GRBs are dense and hot enough to 
allow for the nuclear processes that lead to neutron excess 
(i.e., neutron to proton number density ratio above 1).
The reactions of electron and positron capture on nucleons
and neutron decay must establish nuclear equilibrium. 
These reactions are
\begin{eqnarray}
\label{eq:urca}
p + e^{-} \to n + \nu_{\rm e} \nonumber \\
p + \bar\nu_{\rm e} \to n + e^{+}  \nonumber \\
p + e^{-} + \bar\nu_{e} \to n \nonumber \\
n + e^{+} \to p + \bar\nu_{\rm e} \nonumber \\
n \to p + e^{-} + \bar\nu_{\rm e} \nonumber \\
n + \nu_{\rm e} \to p + e^{-}; 
\end{eqnarray}

The ratio of protons to nucleons must satisfy the balance between 
their number densities and reaction rates:
\begin{eqnarray}
&&n_{\rm p} (\Gamma_{p + e^{-} \to n + \nu_{\rm e}} + 
	\Gamma_{p + \bar\nu_{\rm e} \to n + e^{+}} +
	\Gamma_{p + e^{-} + \bar\nu_{\rm e} \to n}) 
\nonumber \\
&&=n_{\rm n} (\Gamma_{n + e^{+} \to p + \bar\nu_{\rm e}}+ 
	\Gamma_{n \to p + e^{-} + \nu_{\rm e}}+ 
	\Gamma_{n + \nu_{\rm e} \to p + e^{-}})\;.
\end{eqnarray}

The reaction rates are the sum of forward and backward rates and are given
by the following formulae 
\citep{1998PhRvD..58a3009R, 2005ApJ...629..341K}:

\begin{eqnarray}
\Gamma_{p + e^{-} \rightarrow n + \nu_e} 
&=&\frac{1}{2\pi^3}|M|^2 \int_Q^{\infty}dE_e
        E_ep_e(E_e-Q)^2f_e(1-b_ef_{\nu_e}), \nonumber \\
\Gamma_{p + e^{-} \leftarrow n + \nu_e} 
&=&\frac{1}{2\pi^3}|M|^2 \int_Q^{\infty}dE_e
        E_ep_e(E_e-Q)^2(1-f_e)b_ef_{\nu_e}, \nonumber \\
\Gamma_{n+e^{+}\rightarrow p+\bar\nu_e} 
&=&\frac{1}{2\pi^3}|M|^2 \int_{m_e}^{\infty}dE_e
        E_ep_e(E_e+Q)^2f_{e^+}(1-b_ef_{\bar\nu_e}), \nonumber \\
\Gamma_{n+e^{+}\leftarrow p+\bar\nu_e} 
&=&\frac{1}{2\pi^3}|M|^2 \int_{m_e}^{\infty}dE_e
        E_ep_e(E_e+Q)^2(1-f_{e^+})b_ef_{\bar\nu_e}, \nonumber \\
\Gamma_{n\rightarrow p+e^{-}+\bar\nu_e} 
&=&\frac{1}{2\pi^3}|M|^2 \int_{m_e}^QdE_e
        E_ep_e(Q-E_e)^2 (1-f_{e})(1-b_ef_{\bar\nu_e}), \nonumber \\
\Gamma_{n\leftarrow p+e^{-}+\bar\nu_e} 
&=&\frac{1}{2\pi^3}|M|^2 \int_{m_e}^QdE_e
        E_ep_e(Q-E_e)^2 f_{e}b_ef_{\bar\nu_e}. 
\end{eqnarray}

Here $Q=(m_{n}-m_{p})c^{2}$ is the (positive) difference between neutron and proton masses and $|M|^2$ 
is the averaged transition 
rate which depends on the initial and final states of all participating
particles. For nonrelativistic and noninteracting nucleons,
$|M|^{2}=G_{\rm F}^{2}\cos^{2}\theta_{\rm C}(1+3g_{\rm A}^{2})$,
where $G_{\rm F} \simeq 1.436 \times 10^{-49}~{\rm erg}~ 
{\rm cm}^{3}$ is the Fermi weak interaction constant,
$\theta_{\rm C}$ $(\sin\theta_{\rm C}=0.231)$
is the Cabibbo angle, and
$g_{\rm A}=1.26$ is the axial-vector coupling constant.
Here, $f_{e}$ and $f_{\nu_e}$ are the distribution functions for electrons and
neutrinos, respectively. The chemical potential of neutrinos is 
 zero for the neutrino transparent matter. 
When neutrinos are trapped, the factor $b_{e}$ reflects the
percentage of the partially trapped neutrinos.

In addition, two other conditions that need to be satisfied are
the conservation of the baryon number, 
$n_n+n_p=n_b \times X_{\rm nuc}$, and charge neutrality (Yuan 2005);
,
\begin{equation}
n_{\rm e} = n_{e^{-}} - n_{e^{+}} = n_{\rm p}+n_{\rm e}^{0}.
\end{equation}

The above condition includes the formation of the Helium nuclei.
Therefore, the net number of electrons is equal to the number 
of free protons plus the number of protons in Helium, given by
\begin{equation}
n_{\rm e}^{0} = 2 n_{\rm He} = (1-X_{\rm nuc}){n_{\rm b} \over 2}.
\end{equation}

\subsection{Chemical potential of neutrinos}

In the neutrino transparent regime,
the neutrinos are not thermalized, and the chemical potential of neutrinos 
is negligible (see, e.g., Beloborodov 2003).  
In the neutrino opaque regime when neutrinos are totally trapped, the chemical equilibrium condition
yields $\mu_{\rm e} + \mu_{\rm p} = \mu_{\rm n} + \mu_{\nu}$. The chemical 
potential of neutrinos depends on how many neutrinos and anti-neutrinos 
are trapped, assuming that the number densities of the trapped neutrinos and anti-neutrinos 
are the same, where $\mu_{\nu}=0$.

For the intermediate regime of partially trapped neutrinos, 
the Boltzmann equation should be solved. 
To simplify this problem, we use
the "gray body" model with a blocking factor  $b=\sum_{i=e,\mu,\tau} b_i$.
\citep{2003PhRvD..68f3001S,2007ApJ...664.1011J}.
The neutrino distribution function is then
\begin{equation}
\widetilde{f}_{\nu_i}(p)=\frac{b_i}{\exp(pc/kT)+1} =b_if_{\nu_i},\,\,\,(0 \leq b_i \leq 1).
\end{equation}
When neutrinos are
completely trapped, the blocking factor is $b_{i}=1$.

\subsection{Electron fraction and nucleosynthesis}

\begin{figure}
\includegraphics[width=8cm]{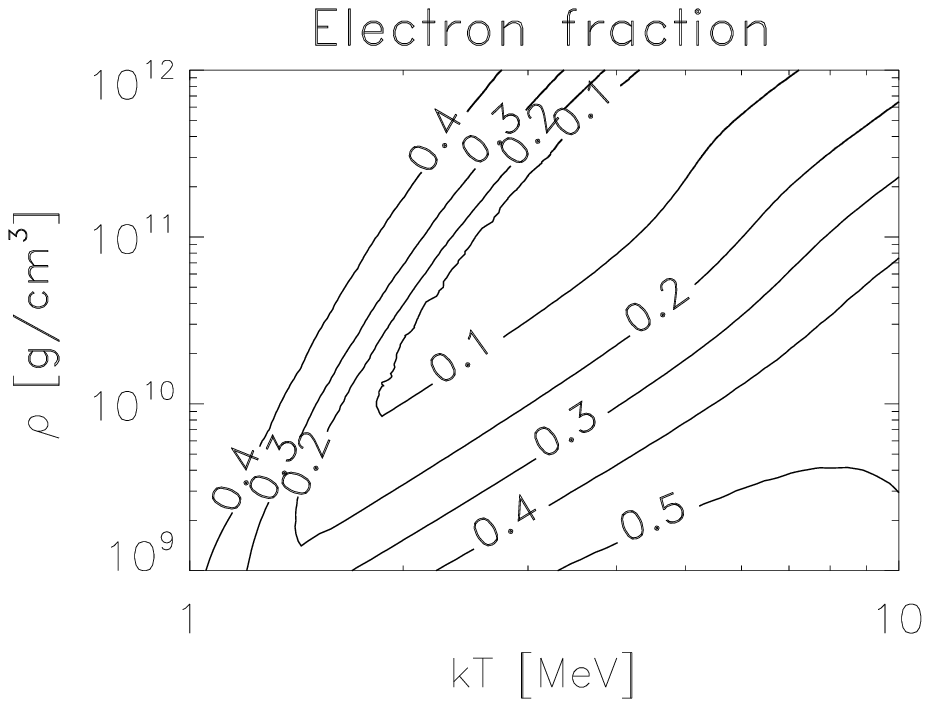}
\caption{Electron-fraction contours in the temperature-density plane.
The matter is partially opaque to neutrinos. Helium nuclei and electron-positron 
pairs are included in the chemical balance.
}
\label{fig:ye1}
\end{figure}

In Figure \ref{fig:ye1}, we show the contours of the constant 
electron fraction on the temperature-density plane.
This value,  calculated as 
\begin{equation}
Y_{e} = (n_{e^{-}}-n_{e^{+}})/n_{b},
\label{eq:ye1}
\end{equation} 
is defined by the
net number of electrons, which is equal to the number density difference between 
electrons and positrons. From the charge neutrality condition, it
is equal to the number of free protons plus the number of electrons in Helium nuclei if they are formed. 
Therefore, the electron fraction is
 modified here by the Helium synthesis process.

The contour of $Y_{e}=0.5$ corresponds to the Helium number density
of $n_{He}=1/4(n_{b}-2n_{p})$. If the neutrons do not dominate over protons,
the Helium density would vanish here.
The top-right region of this figure represents the contours with
$n_{He}<1/4(n_{b}-2n_{p})$, where Helium abundance is smaller and
the electron-positron pairs are produced.

\begin{figure}
\includegraphics[width=8cm]{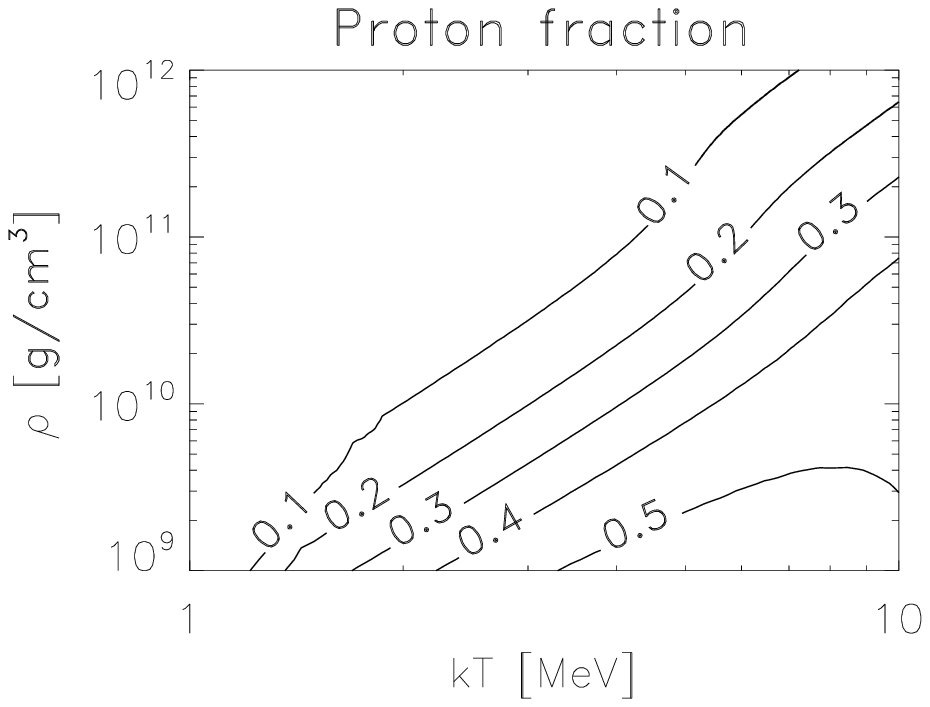}
\caption{Proton-fraction contours in the temperature-density plane. 
The values below 0.5 correspond to the excess of free neutrons over protons.
}
\label{fig:ye2}
\end{figure}

The electron fraction defined above is different from the equilibrium 
value, defined as $Y_{p} = 1/(1+n_{n}/n_{p})$, 
which is given by the ratio of number densities of protons to nucleons.
The reason is due to the presence of electron-positron pairs and Helium nuclei
in the plasma.
This 'proton fraction' is plotted in Figure
\ref{fig:ye2}.
The established chemical balance between the neutronization reactions
leads to the neutron excess, therefore, in the temperature and density range
shown in the Figure; the free proton number density is mostly 
$n_{p}<1/2(n_{p}+n_{n})$.

\section{Accretion hydrodynamics}

We consider the vertically integrated accretion disk around a 
black hole of mass $M$ and a dimensionless spin parameter $a$, 
\citep{2010A&A...509A..55J}.
The surface density of the disk is $\Sigma = H \rho$,
 where
$\rho$ is the baryon density and the height
is given by $H=c_{\rm s}/\Omega_{\rm K}$. Here, the speed of sound is defined by
$c_{\rm s} = \sqrt{P/\rho}$, $\Omega_{\rm K} = c^{3}/(GM(a+(r/r_{g})^{3/2}))$
is the Keplerian angular velocity, $r_{\rm g}=GM/c^{2}$ 
is the gravitational radius,
 and $P$ is the total pressure.
At very high accretion rates, we note that the
disk becomes geometrically 'slim' ($H \sim -0.3- 0.5 r$) in regions,
where neutrino cooling becomes inefficient and advection dominates. 

For the disk viscous stress, we use the standard $\alpha$ viscosity 
prescription of \citet{1973A&A....24..337S},
where the stress 
tensor is proportional to the pressure:
\begin{equation}
\tau_{r\varphi}=-\alpha P,
\end{equation}
and we take a fiducial value of $\alpha=0.1$.

The total pressure is contributed by free nuclei, electron-positron pairs, 
Helium, radiation, and the trapped 
neutrinos.
The 
fraction of  each species is determined by self-consistently solving the
balance of the nuclear reaction rates.
\begin{equation}
P = P_{\rm nucl}+P_{\rm He}+P_{\rm rad}+P_{\nu}\;.
\end{equation} 
where
\begin{equation}
P_{\rm nucl}=P_{\rm e-}+P_{\rm e+}+P_{\rm n}+P_{\rm p}
\end{equation}
with
\begin{equation}
P_{\rm i} = {2 \sqrt{2}\over 3\pi^{2}}
{(m_{i}c^{2})^{4} \over (\hbar c)^{3}}\beta_{i}^{5/2}
\left[F_{3/2}(\eta_{\rm i},\beta_{\rm i})+{1\over 2} \beta_{\rm i}F_{5/2}(\eta_{\rm i},\beta_{\rm i})\right]\;.
\label{eq:pi}
\end{equation}

Here, $F_{\rm k}$ are the Fermi-Dirac integrals on the order $k$, and
$\eta_{\rm e}$, $\eta_{\rm p}$ and $\eta_{\rm n}$ are the reduced chemical
potentials of electrons, protons and neutrons in units of $kT$,
respectively, where $\eta_i = \mu_i/kT$. 
The reduced chemical potential (or degeneracy parameter) of positrons is
$\eta_{\rm e+}=-\eta_{\rm e}-2/\beta_{\rm e}$. The relativity parameters 
of the species $i$ are defined as $\beta_{\rm i}=kT/m_{\rm i}c^{2}$.

The pressure of non-relativistic
and non-degenerate Helium is given by
\begin{equation}
P_{\rm He}= n_{\rm He}kT, 
\end{equation}
where the density is
\begin{equation}
n_{\rm He}={1\over 4}n_{b}(1-X_{\rm nuc})\;,
\end{equation}
and the fraction of free nucleons scales with density and temperature as
\begin{equation}
X_{\rm nuc}=295.5\rho_{10}^{-3/4}T_{11}^{9/8}\exp(-0.8209/T_{11}).
\label{eq:xnuc}
\end{equation}
Here, $T_{11}$ is the temperature in the units of 10$^{11}$ K, 
and $\rho_{10}$ is density
in the units of 10$^{10}$ g cm$^{-3}$ 
\citep{1996ApJ...471..331Q, 1999ApJ...518..356P, 2007ApJ...664.1011J}.

The radiation pressure is negligible in comparison with other terms in the
GRB central engines.
However, when neutrinos are trapped in the disk, 
the neutrino pressure is non-zero. 
Following the treatment of photon transport under the
two-stream approximation \citep{1995ApJ...442..337P, 2002ApJ...579..706D}, 
we have
\begin{eqnarray}
P_{\nu}&=& {7\over8}{\pi^{2}\over 15}{(kT)^{4}\over3(\hbar c)^{3}} 
\sum_{i=e,\mu,\tau} {{1\over 2}(\tau_{a, \nu_{i}}+ \tau_{s}) + {1\over \sqrt 3} \over 
{1\over 2} (\tau_{a, \nu_{i}}+ \tau_{s}) + {1\over \sqrt 3} + {1 \over 3 \tau_{a, \nu_{i}}}} \nonumber \\
&\equiv & {7\over8}{\pi^{2}\over 15}{(kT)^{4}\over3(\hbar c)^{3}}b,
\label{eq:pnu}
\end{eqnarray}
where $\tau_{\rm s}$ is the scattering optical depth due to the neutrino 
scattering on
free neutrons and protons and $\tau_{\rm a, \nu_{e}}$ and $\tau_{\rm a, \nu_{\mu}}$ are 
the absorptive optical depths for electron and muon neutrinos.
The absorption of electron neutrinos is determined by the reactions inverse 
to all their production processes: electron-positron capture on nucleons,
electron-positron pair annihilation, nucleon bremsstrahlung and plasmon decay.
The absorption of muon neutrinos is governed by the rates of pair annihilation and bremsstrahlung reactions, and
the contribution from tau neutrinos is the same as that from muon neutrinos.
The scattering optical depth is the Rosseland mean opacity,as derived for all
the neutrinos from their cross-section of scattering on nucleons \citep{2002ApJ...579..706D}.

Throughout the calculations we adopt fiducial values of parameters: 
black hole mass of $M=3 M_{\odot}$, dimensionless spin $a=0.9$, 
and viscosity parameter $\alpha=0.1$.
We solve the hydrodynamical balance to establish the structure of the accretion disk, by adopting the standard mass, energy,
 and momentum conservation equations.
The viscous heating, $Q^{+}= 3/2 \alpha P \Omega H$, is balanced by the advective cooling, $Q_{adv} = \Sigma v_{r} T dS/dr$, photodissociation of alpha particles (if they form) and the radiative and neutrino cooling, $Q_{\nu}$.
Therefore, we calculate  the stationary disk configuration from
\begin{equation}
F_{\rm tot} = Q^{+}_{\rm visc} = Q^{-}_{\rm adv}+Q^{-}_{\rm
  rad}+Q^{-}_{\nu} + Q_{\rm photo}.
\label{eq:balance}
\end{equation}
On the left hand side, we have the vertically integrated viscous
heating rate per unit area over a half thickness, $H$, which is given by the global parameters of the model; 
that is black hole mass and accretion rate,
\begin{equation} 
F_{\rm tot} = {3 G M \dot M \over 8 \pi r^3} f(r),
\label{eq:ftot} 
\end{equation}
 where $f(r)$ denotes the boundary condition around the spinning Kerr 
black hole \citep{1972ApJ...178..347B,1995ApJ...450..508R}.

In the advective cooling term, the entropy gradient is assumed to be constant with 
radius and be on the order of unity. The entropy is the sum of four components, 
which correspond to the pressure terms
\begin{equation}
S = S_{\rm nucl}+S_{\rm He}+S_{\rm rad}+S_{\nu}\;.
\end{equation}
Here, 
\begin{equation}
S_{\rm nucl}= S_{\rm e^{-}}+S_{\rm e^{+}}+S_{\rm p}+S_{\rm n},
\end{equation}
and the contributions from electrons, positrons, protons and neutrons are
\begin{equation}
\frac{S_{\rm i}}{k} = {1\over kT}(\epsilon_{i}+P_{\rm i})-n_{\rm i}\eta_{\rm i}
\end{equation}
with
\begin{equation}
\epsilon_{i} = {2 \sqrt{2}\over 3\pi^{2}}
{(m_{\rm i}c^{2})^{4} \over (\hbar c)^{3}}\beta_{\rm i}^{5/2}
\left[F_{3/2}(\eta_{\rm i},\beta_{\rm i})+\beta_{\rm i}F_{5/2}(\eta_{\rm i},\beta_{\rm i})\right].
\end{equation}
Here $P_{\rm i}$ is the pressure components of the species,
 $n_{\rm i}$ the number densities, and $\eta_{\rm i}$  the reduced 
chemical potentials.

If the Helium nuclei form, their entropy 
is given by
\begin{equation}
S_{\rm He} = n_{\rm He} \left[{5 \over 2} + {3\over 2} \log (m_{\rm He} {kT\over {(\hbar c)^{2}}}{1\over 2\pi})-\log n_{\rm He}\right]\;,
\end{equation}
for $n_{\rm He}>0$.

The cooling term due to photodisintegration of $\alpha$ particles has a rate of
\begin{equation}
Q_{\rm photo} = q_{\rm photo} H,
\end{equation}   
where
\begin{equation}
q_{\rm photo}= 6.28 \times 10^{28} \rho_{10} v_{r} {dX_{\rm nuc} \over dr},
\label{eq:photodis}
\end{equation}
and $X_{\rm nuc}$ is the mass fraction of free nucleons.

The radiative cooling, $Q^{-}_{\rm rad}=(3 P_{\rm rad} c)/ (4\tau)=(11 \sigma T^{4})/ (4 \kappa \Sigma)$,
 is in practice a negligible term in comparison with other
terms for our assumed global flow parameters.
The neutrino cooling rate is high, if only the neutrinos are not completely 
trapped. 
In the neutrino optically thick disk, their cooling rate is given by
\begin{equation}
Q^{-}_{\nu} = { {7 \over 8} \sigma T^{4} \over 
{3 \over 4}} \sum_{i=e,\mu,\tau} { 1 \over {\tau_{\rm a, \nu_{i}} + \tau_{\rm s} \over 2} 
+ {1 \over \sqrt 3} + 
{1 \over 3\tau_{\rm a, \nu_{i}}}}\;
\label{eq:qnuthick}
\end{equation}
where we include the absorption and scattering optical depths for all three neutrino flavors, following their emission processes,
 as in 
\citet{2007ApJ...664.1011J}.

\section{Results}

\subsection{Mass fraction of free nucleons in the inner disk}

\begin{figure}
\includegraphics[width=8cm]{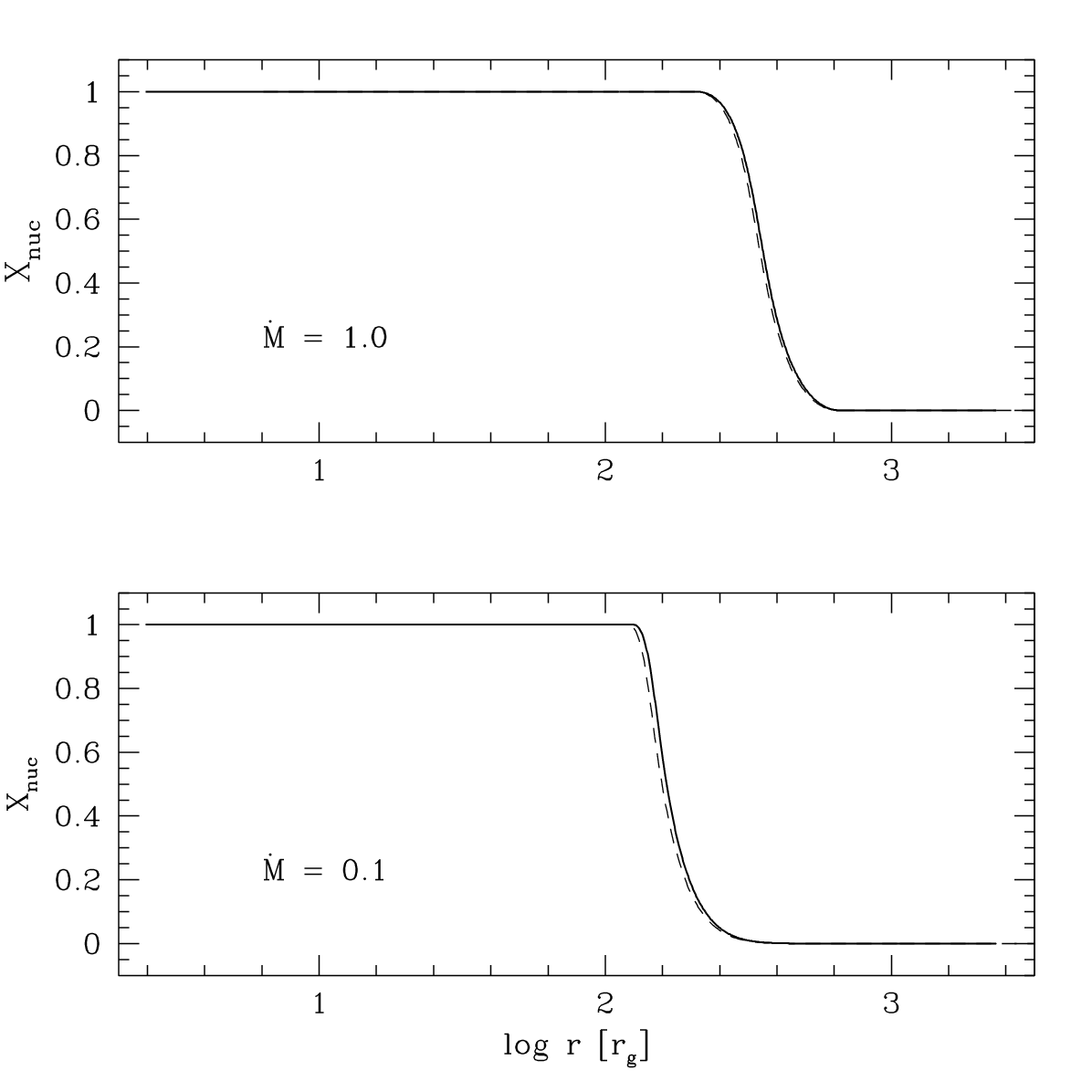}
\caption{Mass fraction of free nucleons as a function of distance
in the accreting disk. The steady-state models were calculated for
$\dot M= 1.0 M_{\odot}/s$ (top panel) and 
$\dot M= 0.1 M_{\odot}/s$ (bottom panel). The radius is 
in units of $r_{g}= G M /c^{2}$ and $M=3 M_{\odot}$.
}
\label{fig:xnuc_disk}
\end{figure}

In Figure \ref{fig:xnuc_disk}, we plot 
the mass fraction of free nucleons in the accretion disk 
that power the central engine of gamma-ray burst.
The Figure shows two values of accretion rate, $\dot M= 0.1 M_{\odot}/s$ and
$\dot M= 1.0 M_{\odot}/s$.
We adopt the black hole mass of $3 M_{\odot}$.

For a given accretion rate, the converged solution for $X_{nuc}$ is not 
sensitive to the black hole spin value and its only effect is on
the location of marginally stable orbit. The disk is located closer to the black hole gravitational radius, $r_{g}=GM/c^{2}$, 
for a large spin value.

Depending on the accretion rate, we find the regions of Helium 
formation in the disk. For a given black hole spin, $a=0.9$, the smaller 
$\dot M=0.1 M_{\odot} s^{-1}$ 
solution results in a constant maximum value of $X_{nuc}=1.0$ below the radius
of about 100 $r_{g}$. For our assumed mass of the black hole it is about 
450 km. Outside this radius, the Helium nuclei start forming. Above $\sim 300 r_{g}$, the mass fraction of free nucleons drops to zero and 
the accreting flow is dominated by Helium.
For a larger value of accretion rate, $\dot M=1.0 M_{\odot} s^{-1}$, this 
outer zone 
dominated by Helium nuclei is shifted outwards to above $\sim 630 r_{g}$; while
below $\sim 250 r_{g}$, there is no Helium nuclei.

\begin{figure}
\includegraphics[width=8cm]{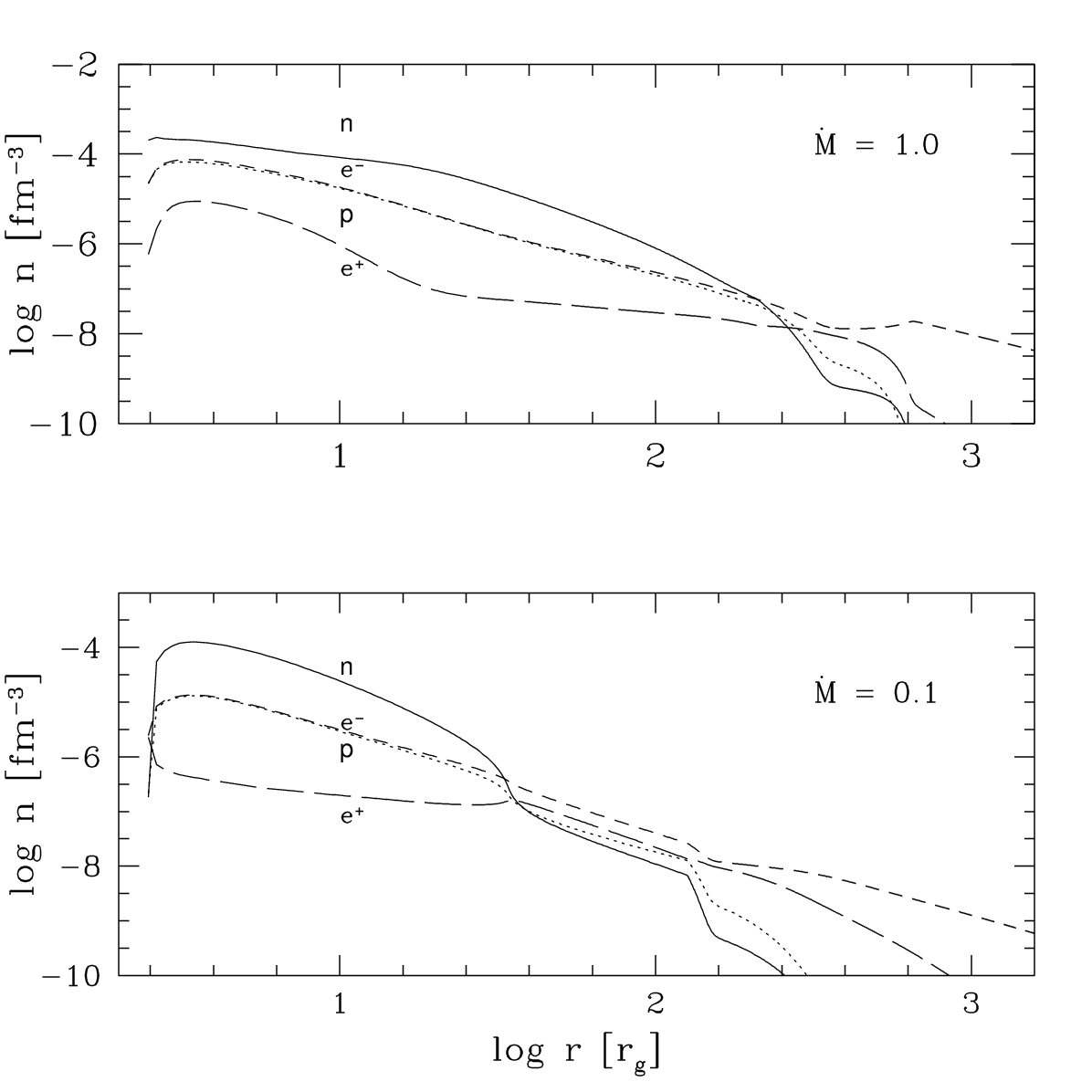}
\caption{Number density of free particles as a function of distance
in the accreting disk, which are plotted for neutrons (solid line), protons (dotted line),
electrons (short dashed line), and positrons (long dashed line).
The steady-state models were calculated for
$\dot M= 1.0 M_{\odot}/s$ (top panel) and 
$\dot M= 0.1 M_{\odot}/s$ (bottom panel). The assumed
 black hole spin parameter is $a=0.9$.
}
\label{fig:chem_disk}
\end{figure}

In Figure \ref{fig:chem_disk}, we show the number densities of free particles
in the disks with two accretion rate values.
Free neutrons are dominant species in a large part of the disk up to
~40 $r_{g}$ for $\dot M= 0.1 M_{\odot}/s$ and up to ~250 $r_{g}$ for 
$\dot M= 1.0 M_{\odot}/s$.
Out from this region, neutrons and protons are synthesizing the Helium nuclei,
and the electron-positron pairs are still abundant, which
keeps the charge neutrality.

For $\dot M= 1.0 M_{\odot}/s$ in the inner $10 r_{g}$, 
the number density of free neutrons is reduced with a flattened radial profile, and the density of protons is
smaller. 
The reason is that  protons 
are more degenerate than neutrons at these 
density and temperature regime.

\begin{figure}
\includegraphics[width=8cm]{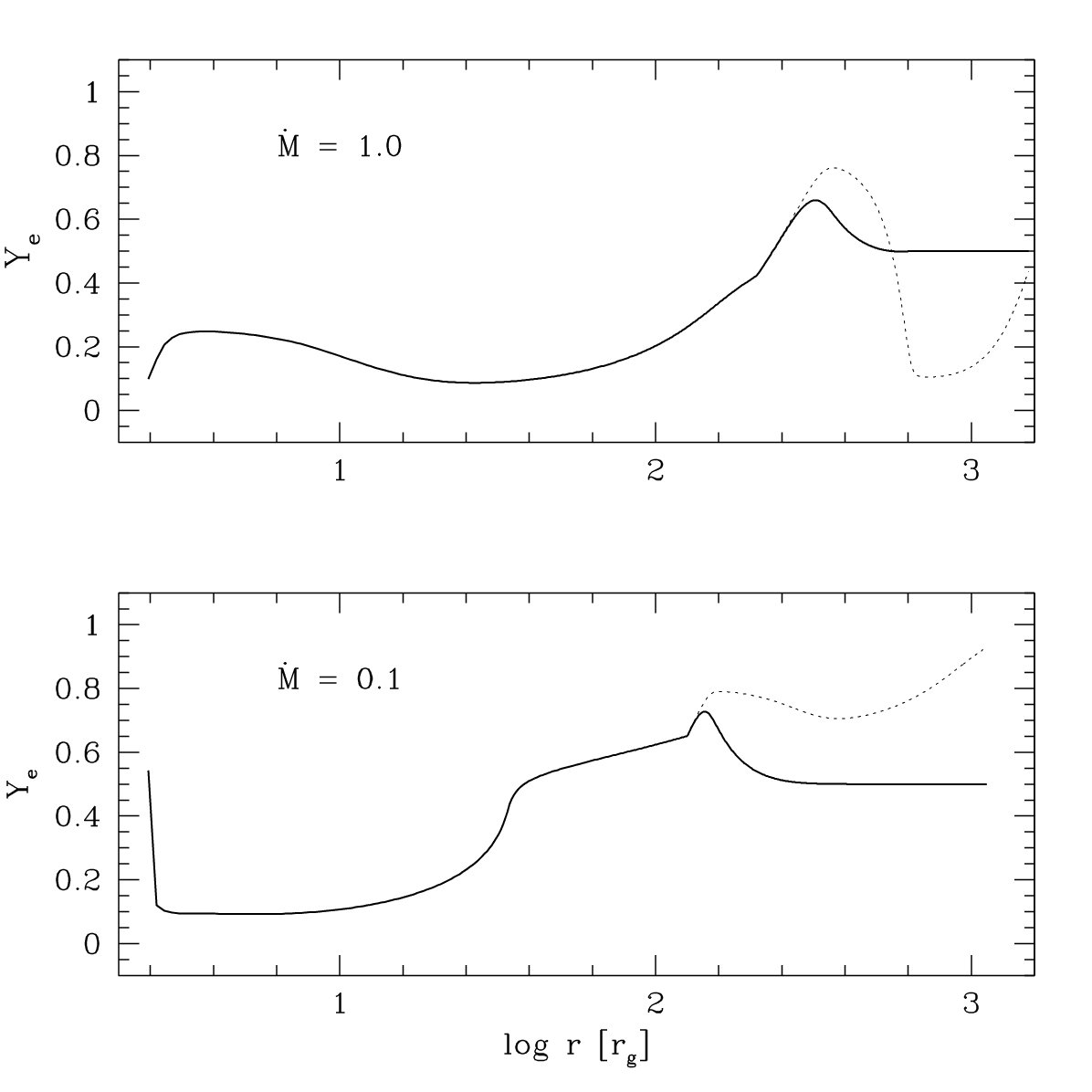}
\caption{Electron fraction as a function of distance
in the accreting disk. The steady state models were calculated for
$\dot M= 1.0 M_{\odot}/s$ (top panel) and 
$\dot M= 0.1 M_{\odot}/s$ (bottom panel). The assumed
 black hole spin parameter is $a=0.9$.
Dashed lines show the corresponding proton fraction.
}
\label{fig:ye_disk}
\end{figure}

In Figure \ref{fig:ye_disk}, we show the electron fraction $Y_{e}$, which is calculated as
the net number of electrons per baryon (Eq. \ref{eq:ye1}).
The dashed line in this figure compares the 'proton fraction',
as derived from the ratio of free neutrons to protons. If the latter is smaller 
than 0.5, the free neutrons dominate over protons, which is the case in the
innermost $50-500$ gravitational radii of the disk, depending on its 
accretion rate. The electron fraction, which is plotted with the solid line,
changes due to the formation of Helium nuclei.
In the outer regions of the disk, the value of electron fraction is
saturated at 0.5, while the number densities of neutrons are
slightly less than protons, but increase with radius. The efficient 
neutronization eventually allows for the maximum Helium abundance at the 
outskirts of the disk.

\subsection{Formation of heavy elements in the outer disk}

The steady-state models were computed for
accretion rates of $\dot M=1.0 M_{\odot}/s$ and $\dot M= 0.1 M_{\odot}/s$ 
with a 
black hole dimensionless spin equal to $a=0.9$. 
In these conditions, the density range in the disk, which is up to 1000 gravitational radii,
is $3\times 10^{11} - 10^{6}$ g cm$^{-3}$ or $5\times 10^{11} - 10^{7}$ g cm$^{-3}$ 
for accretion rates of 0.1 and 1.0 $M_{\odot}$s$^{-1}$, respectively.
The corresponding temperature range is $5\times 10^{10} - 1.5 \times 10^{9}$ K 
and $1.2\times 10^{11} - 2 \times 10^{9}$ K. The innermost temperatures, therefore, 
may reach 5 or 10 MeV, while the regions, where temperature
decreases below 1 MeV, are located in the outer disk parts above 50 or 100 gravitational
radii.

To compute the abundances of heavy elements, we used the
thermonuclear reaction network code 
(http://webnucleo.org).
The computational 
methods are described in detail in \citet{2008ApJ...685L.129S} 
(see also \citealt{Meyer94}, \citealt{Wallerstein97}, and \citealt{HixMeyer}).
The code uses the {\it nuceq} library to compute the nuclear statistical equilibria
established for the thermonuclear fusion reactions.
The abundances are calculated under the constraints of nucleon number conservation
and charge neutrality. We used the correction function to account for
degeneracy of relativistic species here. 
 We used the data
 downloaded from JINA website (http://www.jinaweb.org), prepared for studies of the nuclear masses and 
nuclear partition functions, and for computations of the 
nuclear statistical equilibria. 
The reaction data available on JINA {\it reaclib} online database 
provide the currently best determined reaction rates. 
The nuclide data are merged with reaction data into a network data file, as prepared for
astrophysical applications. The network is 
working well for the temperature ranges about or below 1 MeV, and
our method is appropriate for 
the outer parts of accretion disks in GRBs. For hotter astrophysical plasmas, more advanced
computations could be appropriate \citep{2012IJMPD..2150009K}.

\begin{figure}
\includegraphics[width=8cm]{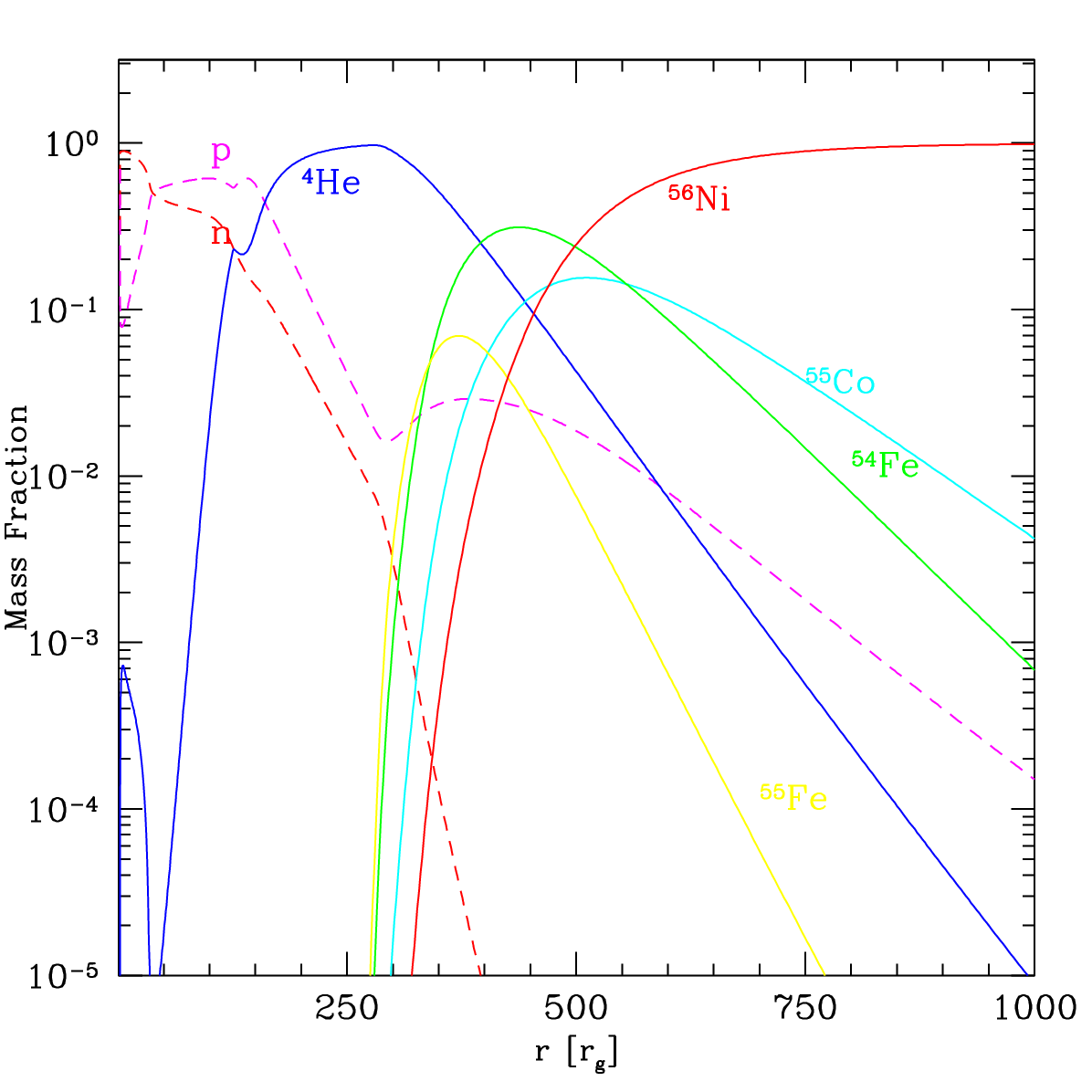}
\includegraphics[width=8cm]{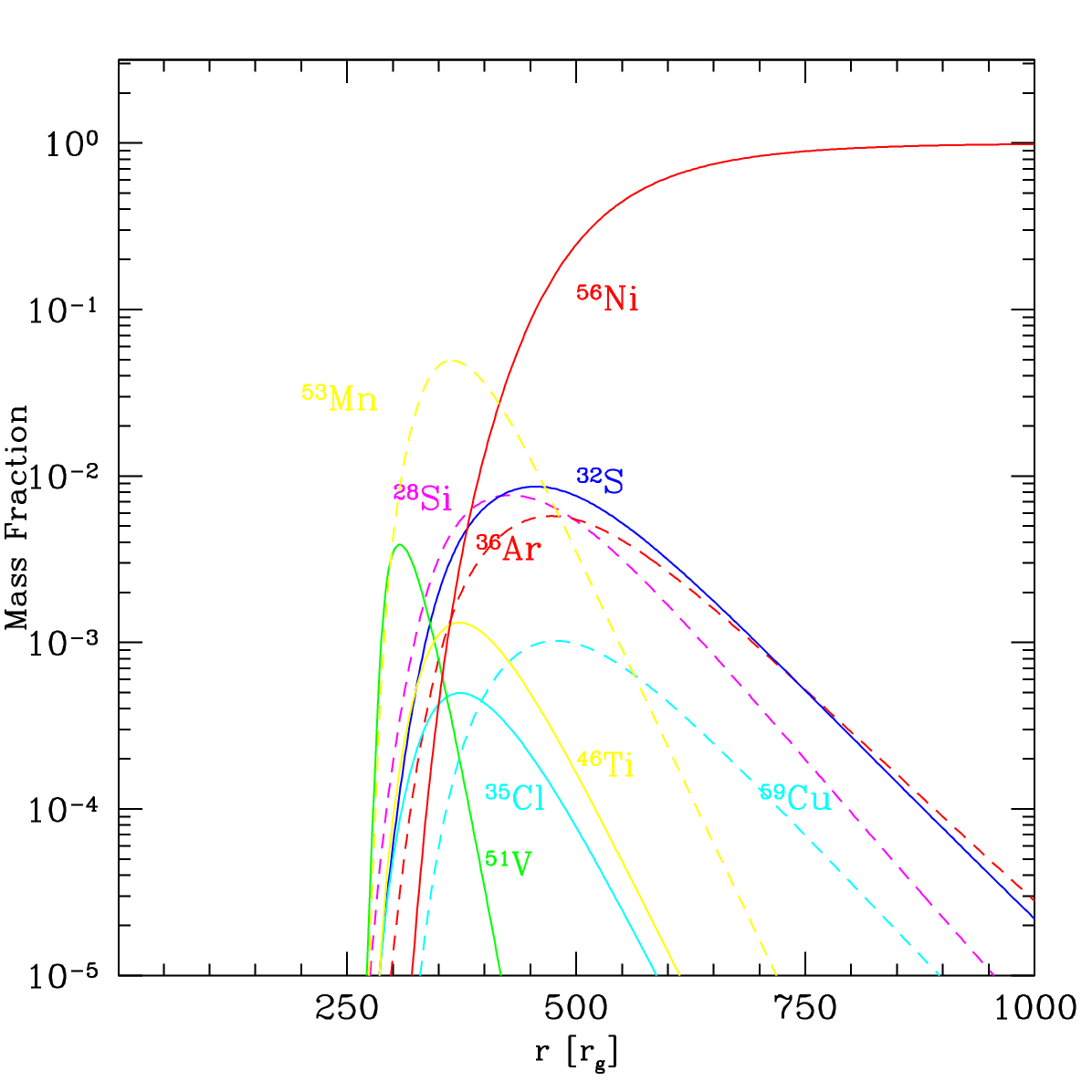}
\caption{Nucleosynthesis of heavy elements in the 
accretion disk. The steady-state model is calculated for
an accretion rate of $\dot M= 0.1 M_{\odot}/s$ and
 a black hole spin  $a=0.9$.
The top panel shows the abundance distribution of 
free protons and neutrons (dashed lines), as well as Helium, and the most abundant isotopes of Nickel, Iron, and Cobalt. 
Bottom panel shows the distribution of the most abundant isotopes of Silicon, 
Sulphur, Clorium, Argonium, Manganium, Titatnium, Vanadium, and Cuprum.
}
\label{fig:nseM21}
\end{figure}

\begin{figure}
\includegraphics[width=8cm]{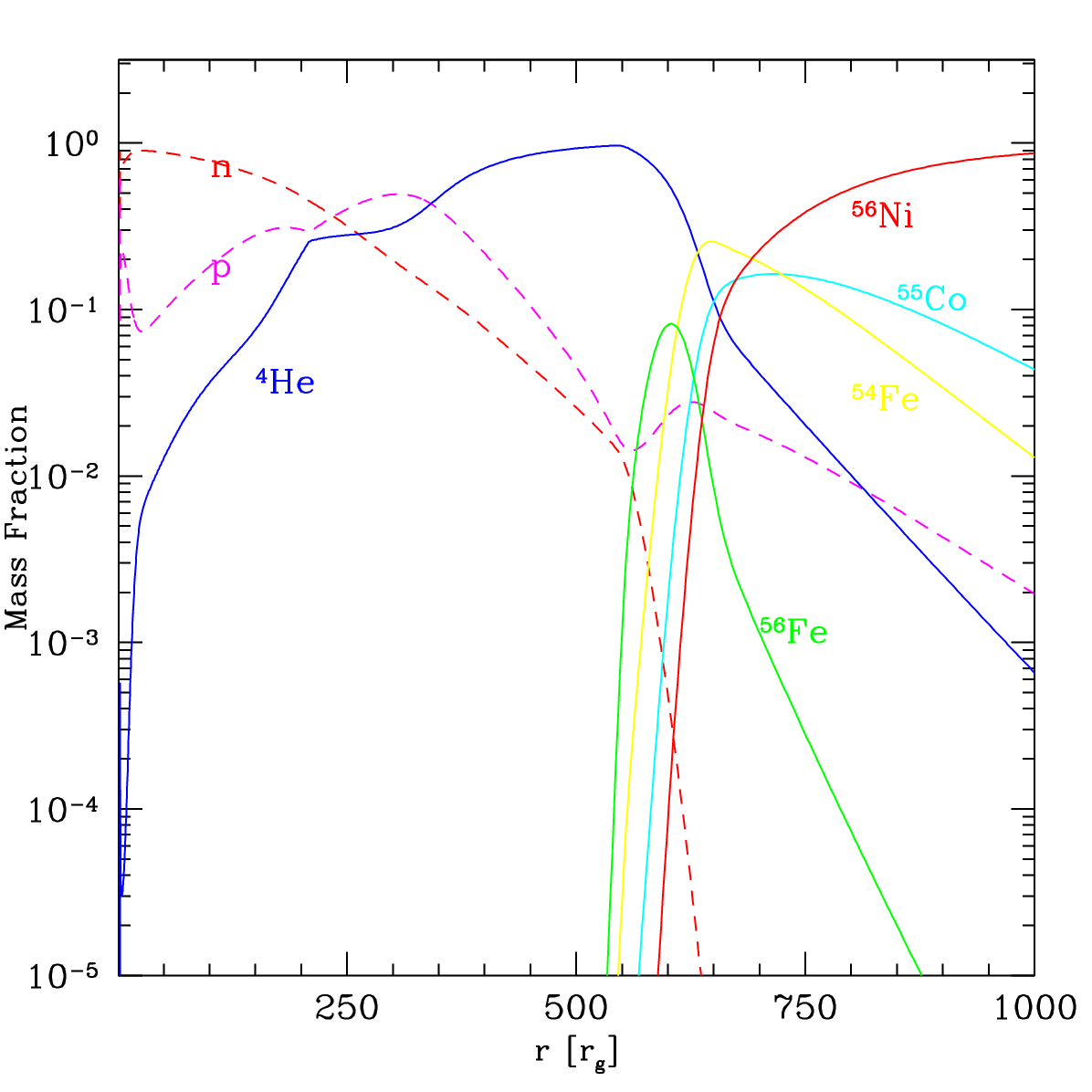}
\includegraphics[width=8cm]{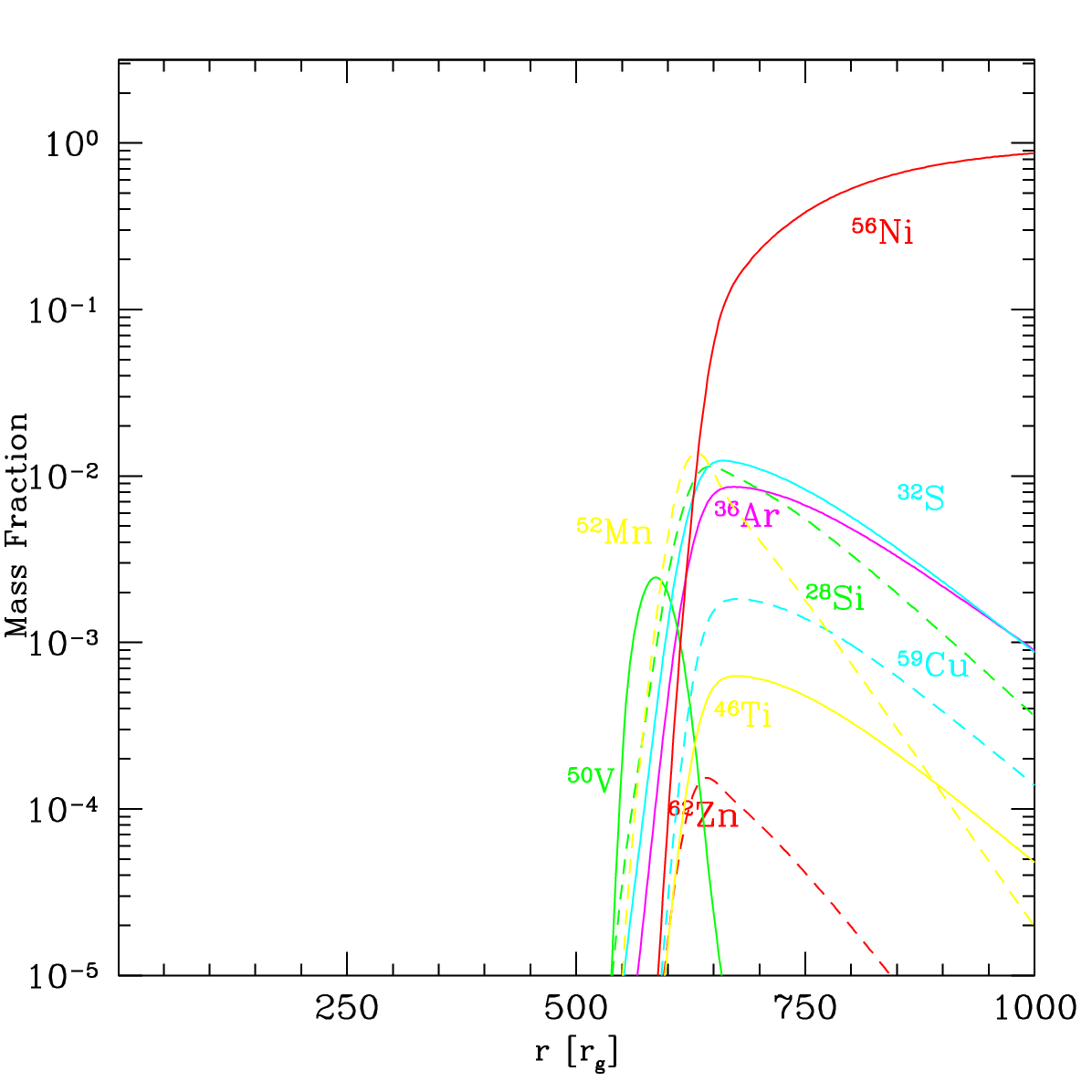}
\caption{Nucleosynthesis of heavy elements in the 
accretion disk. The steady-state model is calculated for
an accretion rate of $\dot M= 1.0 M_{\odot}/s$ and
a black hole spin $a=0.9$.
The top panel shows the abundance distribution of 
free protons and neutrons (dashed lines) as well as Helium, and most abundant isotopes of Nickel, Iron and Cobalt. 
The bottom panel shows the distribution of the most abundant isotopes of 
Argonium, Titatnium, Cuprum, 
and Zinc.
}
\label{fig:nseM22}
\end{figure}

Once we have computed the accretion disk model for the GRB central engine, 
the mass fraction of heavy nuclei was then computed at every radius
of the disk, given
the profiles of  density, temperature and electron fraction.
In Figures \ref{fig:nseM21} and \ref{fig:nseM22}, we show the resulting distributions of the most abundant 
isotopes of heavy elements
synthesized in the accretion disk.

We first checked for these isotopes, whose mass fraction is greater than $10^{-4}$. 
For the accretion rate of 0.1 $M_{\odot}$s$^{-1}$, the abundance of $^{4}He$ is large with a value 
up to 260 $r_{g}$ and then 
decreases
throughout the disk;
there is some fraction of $^{3}He$, Deuterium, and Tritium.
The next abundant isotopes are $^{28}Si$ - $^{30}Si$, $^{31}P$, $^{32}S$ - $^{34}S$, then
$^{35}Cl$, and $^{36}Ar$ - $^{38}Ar$.
Further, synthesized isotopes are $^{39}K$, $^{40}Ca$ - $^{42}Ca$, $^{44}Ti$ - $^{50}Ti$,
$^{47}V$ - $^{52}V$, $^{48}Cr$ - $^{54}Cr$, and $^{51}Mn$ - $^{56}Mn$.
The most abundant Iron isotopes formed in the disk are $^{52}Fe$ through  $^{58}Fe$; Cobalt is formed
with isotopes $^{54}Co$ through  $^{60}Co$, and Nickel isotopes are $^{56}Ni$ through  $^{62}Ni$.
The heaviest most abundant isotopes in our disk are  $^{59}Cu$ through  $^{63}Cu$.
Further, there is a smaller fraction of Zinc, $^{60}Zn$ - $^{64}Zn$, with a mass fraction above $10^{-5}$.
These heavy elements are generally produced outside 300-400 $r_{g}$. Inside this radius,
the disk consists of mainly free neutrons and protons with some fraction of Helium.
The mass fraction of free neutrons is smaller than that of protons, and
free neutrons disappear above $\sim 300 r_{g}$.

For the model with an accretion rate of 1.0 $M_{\odot}$s$^{-1}$,  the abundance of  $^{4}He$ is large, up to
about 600 $r_{g}$ and then  
decreases. There is also some Deuterium and Tritium, while
the next abundant isotopes are $^{28}Si$ - $^{30}Si$, $^{31}P$, 
$^{32}S$ - $^{34}S$, 
$^{35}Cl$, and $^{36}Ar$ - $^{38}Ar$. Next, there is $^{39}K$,
$^{40}Ca$ - $^{42}Ca$, $^{45}Sc$ - $^{47}Sc$, $^{44}Ti$ - $^{50}Ti$,
and $^{47}V$ - $^{52}V$. Synthesized isotopes include $^{48}Cr$ - $^{55}Cr$, 
$^{50}Mn$ - $^{57}Mn$, and $^{52}Fe$ - $^{59}Fe$. 
Further heavy elements are $^{54}Co$ - $^{61}Co$, 
$^{56}Ni$ - $^{63}Ni$, and $^{58}Cu$ - $^{63}Cu$.
The last abundant heavy isotope is $^{62}Zn$, while the abundances of
 $^{65}Ga$ and $^{67}Ga$ are about $10^{-6}$ .
In comparison to the model with small accretion rate presented above,
the conditions in the larger accretion rate disk are such that the mass fraction
of free neutrons
is larger than that of free protons inside $\sim 200 r_{g}$ and comparable
to a proton mass fraction up to $\sim 500 r_{g}$.
In both models, the heavy elements dominate above $\sim 550 r_{g}$.

\section{Outflows}

The outflow of gas from the accretion disk surface may be driven by a
centrifugal force or a magnetic field \citep{2006MNRAS.368.1561M, 2013ApJ...776..105J}.
In either case, the flow can be described by a 
spherical geometry and velocity depending on the distance that computes 
trajectories
of particles \citep{2004ApJ...603..611S}.
The slowly accelerated outflows will produce heavier elements
via the triple-alpha reactions up to Nickel 56, or, if the entropy per baryon is
quite low, the reactions produce nuclei up to Iron peak \citep{1973ApJ...186..601W}.
Also, other heavy nuclei such as Sc, Ti, Zi, and Mo, may be produced 
in the outflows with a moderate abundance \citep{2006ApJ...643.1057S}.

For the accretion rate of $0.1 M_{\odot}/s$, our calculations show the significant proton excess in the disk
 above $\sim 250 r_{g}$. The wind ejected
at this region may therefore provide a substantial abundance of
light elements, Li, Be, and B. 
The high-accretion rate disk, on the other hand,
produces neutron rich outflows and forms heavy nuclei 
via the $r$-process. 
As we show here, 
the outflows ejected from the innermost $100 r_{g}$ in the high-accretion rate disks
 are also significantly neutron rich.
Therefore these neutron-loaded ejecta, which are accelerated via the black hole rotation, 
 feed the collimated jets at a large distance from the central engine.
This has important implications for the observed GRB afterglows, which are
induced by the radiation drag \citep{2008ApJ...676.1130M}
and collisions between 
the proton-rich and neutron-rich shells within the GRB fireball 
\citep{2003ApJ...588..931B}.

\section{Conclusions}

We considered the central engine model for 
gamma-ray bursts, which result from
the massive rotating star collapse or a compact object merger.
The two accretion rates invoked as model parameters refer to these
two progenitor types, which is usually suggested for long and short GRBs.
Our numerical modeling of the accretion flow around a spinning black hole shows that  the
flow is transparent to neutrinos and free species are only mildly degenerate for moderate accretion rates ($\sim 0.1 M_{\odot} s^{-1}$). 
Helium nucleosynthesis occurs at the distance of $\sim 100$ gravitational radii. For large accretion rates above
$\sim 1 M_{\odot} s^{-1}$, the innermost part of the disk is opaque to 
neutrinos. 
Here, the 
 neutrons, protons, and electrons are degenerate. The outflows
from this region should be neutron rich. 
The neutrons, therefore, should be present at large distances within the expanding fireball (i.e. $\sim 10^{17}$ cm).
Their large kinetic energy affects the dynamics of the expanding fireball 
and leads to interaction with the circumburst medium \citep{2013ApJ...765..125L}.

Our results are in line with those obtained by other groups, who are
working on nucleosynthesis models of GRB engines. In particular,
\citet{2004ApJ...614..847F} also compute the nuclear synthesis of elements in the accretion disk up to 1000 gravitational radii, and they find more distant layers
of dominant Oxygen, Silicon, and Calcium with their abundances enhanced by the
$\alpha$-capture process. With a mass fraction about $\sim 1$, these elements
may be present in large collapsar disks. 
However, the disk is probably not larger in most progenitors.
Nevertheless, we obtain similar results for the isotopes synthesized above 
$\sim 100 r_{g}$ (note that we define $r_{g}=G M/c^{2}$), and we also find a trend of shifting the layers outward with an 
increasing accretion rate. We estimate the transition radius for the subsequent layers to be  
$r_{tr} \sim \dot M^{-0.28}$.
At the outer parts of the disk, \citet{2004ApJ...614..847F} studied the synthesis of light elements such as $^{16}O$, $^{28}Si$, 
and $^{40}Ca$, which has mass fractions
above 0.0001. We find the mass fraction of $^{16}O$ to be less than $10^{-4}$ inside 1000 gravitational radii, while they are about
0.01 at their peak, at around few hundred $r_{g}$, for the other two elements . 
For the heavier elements above the Iron peak, their yields were
computed by \citet{2006ApJ...643.1057S}. Our reaction network results
give the isotopes of $^{74}Se$, $^{80}Kr$, and $^{84}Sr$ with the abundances 
of about $10^{-10}$, $10^{-12}$ and $10^{-14}$, 
respectively, in the case of smaller accretion rate disk. We also found
isotopes of $^{84}Rb$ and $^{90}Zr$ with mass fractions of $10^{-12}$ and $10^{-14}$, respectively, for a 
higher accretion rate disk model.

The main difference among our work, those of \citet{2004ApJ...614..847F} and those of
\citet{2004ApJ...603..611S} results from
our model of the accretion disk and detailed 
treatment of its microphysics, which follows the 
equation of state discussed in detail in \citet{2007ApJ...664.1011J}.
In the resulting structure of the disk below $\sim 10$ $r_{g}$,
we have more neutron-rich material, and the ratio of free neutrons to protons exceeds
$n_{n}/n_{p}=10$ at the inner disk region for $\dot M = 0.1 M_{\odot}/s$. 
Outward of $\sim 50$ $r_{g}$, the disk is proton rich.
For a higher accretion rate, this neutron rich region is shifted outward to  
$ > 100 r_{g}$, while we have $n_{n}/n_{p} \approx 3$ at the inner edge . 
The electron fraction profiles in our models have the same qualitative trends
as the other authors have found and saturate at $Y_{e}=0.5$ at the outer disk edge, 
while they decrease inward to $Y_{e} \sim 0.1$ due to electron capture. Overall, our profiles
are below those found by \citet{2004ApJ...614..847F}. 
We also find a small bump with $Y_{e}\sim 0.7$ around the radius 160-250 $r_{g}$, 
depending on accretion rate, located where most of the Helium is synthesized.
Another small bump, where the electron fraction locally rises to $Y_{e} \sim 0.2$,
is found for high-accretion rate of $\dot M = 1.0 M_{\odot}/s$.

The effects of photodissociation of Helium 
might have an important effect on its stability properties
\citep{2007ApJ...664.1011J}. 
This destabilizing process is not suppressed by forces, such as by 
a magnetic torque that is exerted on the disk by a rotating black hole 
\citep{2009ApJ...700.1970L, 2010A&A...509A..55J}.

The heavy elements, up to $^{56}Ni$, are synthesized efficiently in the 
outer parts of the disk.
These nucleosynthesis processes occur at above 
$\sim 250 r_{g}$ or  $\sim 500 r_{g}$, depending on the accretion rate. These
elements, such as Silicon, Phosphorus, Sulphur, Calcium, Argonium, Potassium, and the various isotopes,
Titanium, Vanadium, Chromium, Manganese, Iron, and Cobalt, are present in wind outflows ejected from the surface 
of accretion disks. These elements should be observable in the GRB spectra \citep{2003A&A...403..463R},
 and help constrain the properties of the central engine in the 
GRB explosions (see, e.g., \citealt{2013ApJ...778...18M}).

Moreover, we found 
non-negligible abundances of elements above Nickel, such as 
isotopes of Copper, Zinc, Gallium, and Germanium, which 
could be found in the afterglows of short GRBs. The different properties
of central engines in the two classes of bursts that determine the 
nucleosynthesis should therefore also be accounted 
for in the statistical studies of the observed phenomena  
\citep{2006ApJ...642..354Z, 2008ApJ...689.1161G}.

The radioactive decay of certain isotopes should be detectable via 
the emission lines observed by X-ray satellites. These lines, such as those found in
the decay of $^{44}Ti$ to $^{40}Ca$ with emission of hard X-ray photons
 at 68 and 78 keV, have been detected by NuSTAR in the case of supernova remnants.
The energy band of this instrument (3-80 keV) should be appropriate to find
 the X-ray signatures of other elements synthesized in
the accretion disks in GRB central engines, like
the radioactive isotopes of Cuprum, Zinc, Gallium, Cromium, and Cobalt.
The heavy elements, which were once ejected with a wind outflow from the engine, might also
give their imprints in the absorption lines of the GRB optical spectra
\citep{2006Natur.440..184K}.

\section*{Acknowledgments}
We thank Irek Janiuk for discussions and 
help in parallelization of our code.
We also thank Carole Mundell and Greg Madejski, and
the anonymous referee for helpful comments.
This work was supported in part by the grant NN 203 512 638, 
 from the
Polish National Science Center. We finally acknowledge inspiration and 
support by the COST Action MP0905.

\bibliographystyle{aa} 
\bibliography{grb_nucl} 

\end{document}